\documentclass[letterpaper, 10 pt, conference]{ieeeconf}  

\usepackage{graphicx}
\usepackage{tabularx,booktabs}
\usepackage{color}
\usepackage{multicol}
\usepackage{amsmath,amssymb}
\usepackage{amsfonts}
\usepackage{textcomp}
\usepackage{psfrag}
\usepackage{multimedia}
\usepackage{fancybox}
\usepackage{algorithm}
\usepackage{algorithmic}
\usepackage{varioref}
\usepackage{import}
\usepackage{framed,xcolor}
\usepackage{color}
\usepackage{epstopdf}
\usepackage{float}
\usepackage{etoolbox}
\usepackage{soul}

\IEEEoverridecommandlockouts                              

\definecolor{seb}{rgb}{0.8,1,0.8}

\definecolor{hos}{rgb}{0.91,0.84,0.42}

\newcommand{\vect}[1]{\ensuremath{\boldsymbol{\mathrm{#1}}}}

\newtheorem{Lemma}{Lemma}

\title{\LARGE \bf
Backstepping-based Integral Sliding Mode Control with Time Delay Estimation for Autonomous Underwater Vehicles 
}

\author{Hossein Nejatbakhsh Esfahani, Behdad Aminian, Esten Ingar Grøtli, S\'ebastien Gros
	\thanks{Authors are with Department of Engineering Cybernetics, Norwegian
		University of Science and Technology (NTNU).
		{\tt\small \{hossein.n.esfahani, behdad.aminian, sebastien.gros\}@ntnu.no and estenIngar.grotli@sintef.no%
		}
}}

\begin{document}

\maketitle
\begin{abstract}
    The aim of this paper is to propose a high performance control approach for trajectory tracking of Autonomous Underwater Vehicles (AUVs). However, the controller performance can be affected by the unknown perturbations including model uncertainties and external time-varying disturbances in an undersea environment. To address this problem, a Backstepping-based Integral Sliding Mode Control with Time Delay Estimation (BS-ISMC-TDE) is designed. To improve the performance of a conventional backstepping control algorithm, an Integral Sliding Mode Control (ISMC) approach is adopted in the backstepping design to attenuate the steady-state error. Moreover, an adaptive Time Delay Estimation (TDE) strategy is proposed to provide an estimation of perturbations by observing the inputs and the states of the AUV one step into the past without an exact knowledge of the dynamics and the upper bound of uncertainties. From the simulation results, it is shown that the proposed control approach using both adaptive and conventional TDE parts outperforms a Backstepping-based Integral Sliding Mode Control (BS-ISMC).  
\end{abstract}
\section{Introduction}
In recent years, underwater vehicles have been extensively
developed because of needs in marine industries and their applications such as nondestructive test of the marine structures or underwater oil/gas pipeline and ocean exploration. One of the main reasons that the autonomous marine operation systems in particular the Autonomous Underwater Vehicles (AUVs) are getting more attention is potentially dangerous environment and precise tasks which are difficult and time consuming to accomplish manually \cite{auv-app}.

\par However, the control of the AUVs in a fully autonomous mode is a major challenging problem since this platform includes a coupled and highly nonlinear dynamic. There are severe external time-varying disturbances that affect the tracking performance for such an autonomous platform. Furthermore, the uncertainties raised from the hydrodynamics forces, the added inertia matrix, the added Coriolis and centripetal matrix, and the hydrodynamic damping terms should be considered and tackled.
\par In \cite{AUV1}, an adaptive fuzzy Sliding Mode Control (SMC) approach was proposed while a neural network compensator was adopted to cope with the uncertainties. A second-order sliding-mode control algorithm for suppressing the parameter uncertainties and external disturbances was proposed in \cite{AUV2}. To eliminate the effects of uncertainties and chattering phenomenon, a hybrid robust control scheme including the SMC and $H^\infty$ was proposed in \cite{AUV3}. A robust model predictive control of underwater-vehicle manipulator systems was proposed in \cite{UVMS}. In this robust control scheme, an uncertainty compensator based on fuzzy logic was designed. In the context of backstepping (BS) algorithm as another well-known robust control approach for AUVs, some works were conducted in \cite{AUV4} and \cite{AUV5}. Although the BS is a well-known, simple and useful control approach for the nonlinear systems, the algorithm is not enough robust against both matched and mismatched uncertainties. Therefore, people usually combine a BS with adaptive control approaches or SMC method. Besides, BS requires a full knowledge of the dynamic and the derivatives of the states, which it is mostly difficult to have a quite accurate model for an AUV. 
\par To improve the trajectory tracking accuracy of AUV, authors in \cite{AUV6} proposed to leverage the properties of two robust control schemes as a combined backstepping-based terminal sliding mode controller. In the SMC-based backstepping control algorithms, the SMC law is adopted in the first step of the backstepping design to attenuate the steady-state error and to improve the robustness in presence of mismatched and state dependant uncertainties.
\par 
The advantages of using SMC include fast response and convergence, promising transient performance and acceptable robustness. The Time Delay Estimation (TDE) concept is a key part of Time Delay Control (TDC) approach so as to take advantage of its model-free nature \cite{new_TDE1,new_TDE2,tdc1}. Indeed, the TDE provides an estimation of perturbations by observing the inputs and the states of the dynamical model one step into the past without an exact knowledge of the dynamics and the upper bound of uncertainties. To address the properties of the SMC and TDC in a unified controller, researchers proposed to combine these robust control schemes in \cite{NEJATBAKHSHESFAHANI201597}, \cite{ESFAHANI2019106526}. 
\par To improve the tracking accuracy for motion control of AUVs under external time-varying disturbances and uncertainties, we propose to leverage three robust control approaches including: Backstepping (BS), Integral Sliding Mode Control (ISMC) and Time Delay Control (TDC) in order to provide a high performance and sufficiently robust control scheme. We also propose a gain adaptation strategy for the TDE part to tune a crucial matrix ($\bar M$) in the context of TDC algorithms.
\par This paper is organized as follows: In Section II, a four degrees of freedom AUV is modeled and described. Section III, the proposed control approach and a stability analysis based on Lyapunov theory are presented. The effectiveness of the proposed robust control scheme for the trajectory tracking scenarios is investigated and illustrated in Section IV. Finally, conclusions are drawn in Section V.

\section{AUV Model}
\begin{figure}[htbp!]
		\centering
		\includegraphics[width=.9\linewidth]{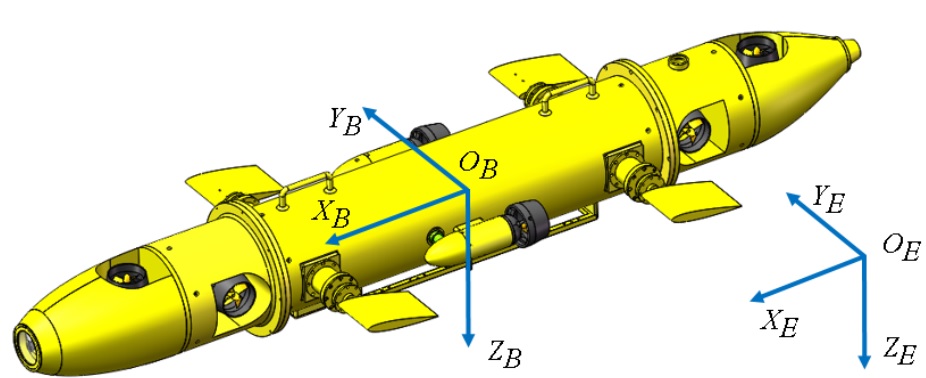}
		\caption{AUV model}
		\label{AUV}
	\end{figure}

The AUV platform is shown in Figure~\ref{AUV}. The dynamics can be described in two reference frames: the Body-fixed frame and the Earth-fixed frame.
Note that due to the effect of the buoyancy force, the pitch and roll motions can be  assumed to be intrinsically stable. However, one can assume that there is higher level controller as autopilot to track the desired roll and pitch angles. Therefore, the 6-DOF platform can be reduced to the fully actuated 4-DOF model with the following nonlinear equations:
\begin{subequations}\label{eq:model_4dof}
		\begin{align}
			&\dot{\vect\eta}=J(\vect\eta)\vect v,\\
			&M\dot{\vect v}+C(\vect v)\vect v+D(\vect v)\vect v+\text{g}(\vect\eta)=\vect\tau+\vect\tau_d
		\end{align}
\end{subequations}
where $\vect\eta=\left[x,y,z,\psi\right]^\top$ is the position-orientation (yaw angle) vector and $\vect v=\left[u,v,w,r\right]^\top$ represents the velocity vector for the surge, sway, heave and yaw motions, respectively. $M, C(\vect v), D(\vect v)\in\mathbb{R}^{4\times4}$ are the inertia matrix (including the effects of added mass), Coriolis-centripetal matrix and the drag force matrix (hydrodynamic damping), respectively. The vector of forces and moments induced by the gravity and buoyancy is labeled by $\text{g}\in\mathbb{R}^{4\times1}$. The control inputs and the external time-varying disturbances (winds, waves, and ocean currents) are labeled as $\vect\tau,\vect\tau_d\in\mathbb{R}^{4\times1}$, respectively. The transformation matrix between the reference frames can be represented in Euler angles as follows:
\begin{align}
    J(\vect\eta)=\begin{bmatrix}
cos(\psi) &-sin(\psi)& 0 &0 \\ 
sin(\psi) &cos(\psi)  &0  &0 \\ 
0 &0  &1  &0 \\ 
0 &0  &0  &1 
\end{bmatrix}
\end{align}
Let $M=diag\left\{ m_{11},m_{22},m_{33},m_{44}\right\}$ be the inertia matrix. Then, the diagonal terms are $m_{11}=m-X_{\dot u}$, $m_{22}=m-Y_{\dot v}$ , $m_{33}=m-Z_{\dot w}$, $m_{44}=I_z-N_{\dot r}$. $m$ is the total mass of AUV and $I_z$ is the moment of inertia about yaw motion. The corresponding hydrodynamic coefficients are labeled by $X_{\star }$,  $Y_{\star }$,  $Z_{\star }$,  $N_{\star }$. Then the Coriolis–centripetal matrix is expressed as
\begin{align}
    C(\vect v)=\begin{bmatrix}
0 &0& 0 &-(m-Y_{\dot v})v \\ 
0 &0  &0  &(m-X_{\dot u})u \\ 
0 &0  &0  &0 \\ 
(m-Y_{\dot v})v &-(m-X_{\dot u})u  &0  &0 
\end{bmatrix}
\end{align}
Let $D(\vect v)=diag\left\{ d_{11},d_{22},d_{33},d_{44}\right\}$ be the damping matrix. Then, the diagonal terms are $d_{11}=-X_u-X_{|u|u}|u|$, $d_{22}=-Y_v-Y_{|v|v}|v|$ , $d_{33}=-Z_w-Z_{|w|w}|w|$, $d_{44}=-N_r-N_{|r|r}|r|$. The gravity-buoyancy vector is specified as follows:
\begin{align}
    \text{g}=\begin{bmatrix}
0\\ 
0\\ 
-(G-B)\\ 
0
\end{bmatrix}
\end{align}
where $G,B$ are the gravity and buoyancy forces, respectively. One can rewrite \eqref{eq:model_4dof} as the following equation:
\begin{align}
    M(\vect\eta)\ddot{\vect\eta}+C(\dot{\vect\eta},\vect\eta)\dot{\vect\eta}+D(\dot{\vect\eta},\vect\eta)\dot{\vect\eta}+\text{g}(\vect\eta)=\vect\tau+\vect\tau_d
\end{align}
where
\begin{subequations}
		\begin{align}
			&M(\vect\eta)=J^{-T}MJ^{-1}(\vect\eta),\\
			&C(\dot{\vect\eta},\vect\eta)=J^{-T}\left[C(\vect v)-MJ^{-1}(\vect\eta)\dot J(\vect\eta)\right]J^{-1}(\vect\eta),\\
			&D(\dot{\vect\eta},\vect\eta)=J^{-T}D(\vect v)J^{-1}(\vect\eta),\\
			&\bar{\vect\tau}=J^{-T}\vect\tau,\\
			&\bar{\vect\tau}_d=J^{-T}\vect\tau_d
		\end{align}
\end{subequations}
The above mentioned dynamics include the uncertain parameters practically. Let $\Delta M,\Delta C,\Delta D,\Delta \text{g}$ be the uncertain parts for each parameter matrix or vector. Then the entire model of AUV can be rewritten as follows:
\begin{align}\label{eq:main_model}
    M(\vect\eta)\ddot{\vect\eta}+C(\dot{\vect\eta},\vect\eta)\dot{\vect\eta}+D(\dot{\vect\eta},\vect\eta)\dot{\vect\eta}+\text{g}(\vect\eta)=\bar{\vect\tau}+\vect p
\end{align}
where, the perturbation $\vect p$ involves the external time-varying disturbances $\vect\tau_d$ and model uncertainties $\delta$ and then:
\begin{align}\label{eq:uncertainty}
   \vect p=&\bar{\vect\tau}_d-J^{-T}\Delta MJ^{-1}(\vect\eta)\ddot{\vect\eta}\\\nonumber
   &-J^{-T}\left[\Delta C(\vect v)-\Delta MJ^{-1}(\vect\eta)\dot J(\vect\eta)\right]J^{-1}(\vect\eta)\dot{\vect\eta}\\\nonumber
   &-J^{-T}\Delta D(\vect v)J^{-1}(\vect\eta)-\Delta\text{g}(\vect\eta)=\bar{\vect\tau}_d-\delta
\end{align}
Let us define $\vect x_1=\vect\eta$ and $\vect x_2=\dot{\vect\eta}$ as state variables and then the concatenated state vector is $\vect x=\left[\vect x_1,\vect x_2\right]^\top$. Then, the system \eqref{eq:main_model} can be described as the following state-space model:
\begin{align}\label{eq:model}
\left\{\begin{matrix}
\dot{\vect x}_1=\vect x_2\\ 
\dot{\vect x}_2=\vect f(\vect x)+\vect u+ \tilde{\vect d}\\ 
\vect y=\vect x_1
\end{matrix}\right.
\end{align}
where
\begin{subequations}
		\begin{align}
			&\vect f(\vect x)=M^{-1}(\vect x_1)\left(-C(\vect x_1,\vect x_2)\vect x_2-D(\vect x_1,\vect x_2)\vect x_2-\text{g}\right),\\
			&\vect u=M^{-1}(\vect x_1)\bar{\vect\tau},\\
			&\tilde{\vect d}=M^{-1}(\vect x_1)\vect p
		\end{align}
\end{subequations}
\section{Control Approach}
In this section, we first propose to leverage the properties of both robust control approaches Backstepping and ISMC as a Backstepping-based Integral Sliding Mode (BS-ISMC) and then enhance the performance of BS-ISMC with considering a TDE part as a compensator to cope with the lumped uncertainties including the unknown uncertain dynamic in the context of the robust control design.  
\subsection{Backstepping-based Integral Sliding Mode}
Let us define the tracking error and its time derivative as follows:
\begin{subequations}
\begin{align}\label{eq:error}
    &\vect e=\vect\eta-\vect\eta_d,\\
    &\dot{\vect e}=\dot{\vect\eta}-\dot{\vect\eta}_d
\end{align}
\end{subequations}
As the first step, the following Lyapunov function is used to ensure the stability of the system:
\begin{subequations}
\begin{align}
    &V_0=\frac{1}{2}\vect e^\top\vect e,\\
    &\dot V_0=\vect e^\top\dot{\vect e}=\vect e^\top(\vect x_2-\dot{\vect\eta}_d)
\end{align}
\end{subequations}
As an step in the backstepping design, the state variable $\vect x_2=\dot{\vect\eta}$ in the state-space model \eqref{eq:model} can be interpreted as a virtual control signal. Then, using the virtual control $\vect x_2=\alpha_0(\vect e)=-k_1\vect e+\dot{\vect\eta}_d$, the origin of $\dot{\vect e}=-k_1\vect e$ is globally asymptotically stable since $\dot V_0=-k_1\vect e^\top\vect e<0,\quad \forall\left(\vect e\neq 0\right)\in\mathcal{R}$ where $k_1$ is a positive diagonal matrix. An integral action can be added to the virtual control as follows:
\begin{align}
    \vect x_2=\alpha(\vect e)=-k_1\vect e-k_2\int_{0}^{t}\vect e(\tau)d\tau+\dot{\vect\eta}_d
\end{align}
where $k_2$ is also a positive diagonal matrix. Then, the time derivative of tracking error can be modified as follows:
\begin{align}\label{eq:de}
    \dot{\vect e}=-k_1\vect e-k_2\int_{0}^{t}\vect e(\tau)d\tau
\end{align}
The Lyapunov function addressing the integral action can be expressed as follows:
\begin{align}
    V_1=\frac{1}{2}\vect e^\top\vect e+\frac{1}{2}k_2\left(\int_{0}^{t}\vect e(\tau)d\tau\right)\left(\int_{0}^{t}\vect e(\tau)d\tau\right)^\top
\end{align}
The time derivative of the above function reads as:
\begin{align}\label{eq:dv1}
    \dot V_1=\vect e^\top\dot{\vect e}+k_2\left(\int_{0}^{t}\vect e(\tau)d\tau\right)\vect e^\top
\end{align}
Then, substituting \eqref{eq:de} to \eqref{eq:dv1}, the derivative of Lyapunov function is delivered as a negative semi-definite $-k_1\vect e^\top\vect e\leq0$ and then the system augmented by an integral action is asymptotically stable in the first stage of the backstepping design. The sliding mode surface based on the deviation of the proposed virtual control form its desired signal in the backstepping control scheme can be defined as follows:
\begin{align}\label{eq:surface}
    \vect\sigma=\dot{\vect\eta}-\alpha(\vect e)=\dot{\vect\eta}+k_1\vect e+k_2\int_{0}^{t}\vect e(\tau)d\tau-\dot{\vect\eta}_d
\end{align}
Considering the tracking error in \eqref{eq:error} and the above sliding surface, the time derivative of tracking error can be obtained as follows:
\begin{align}\label{eq:de2}
    \dot{\vect e}=\vect\sigma-k_1\vect e-k_2\int_{0}^{t}\vect e(\tau)d\tau
\end{align}
Let us define a composite Lyapunov function as:
\begin{align}\label{eq:V}
    V=V_1+\frac{1}{2}\vect\sigma^\top\vect\sigma
\end{align}
considering equations \eqref{eq:dv1}, \eqref{eq:de2} and \eqref{eq:V}:
\begin{align}
    \dot V&=\vect e^\top\dot{\vect e}+k_2\vect e^\top\int_{0}^{t}\vect e(\tau)d\tau+\vect\sigma^\top\dot{\vect\sigma}\nonumber\\
    &=\vect e^\top\left(\vect\sigma-k_1\vect e-k_2\int_{0}^{t}\vect e(\tau)d\tau\right)+k_2\vect e^\top\int_{0}^{t}\vect e(\tau)d\tau\nonumber\\
    &+\vect\sigma^\top\left(\ddot{\vect e}+k_1\dot{\vect e}+k_2\vect e\right)
\end{align}
Substituting $\ddot{\vect e}=\ddot{\vect\eta}-\ddot{\vect\eta}_d$ and \eqref{eq:model} to the above equation:
\begin{align}
    \dot V&=\vect e^\top\left(\vect\sigma-k_1\vect e-k_2\int_{0}^{t}\vect e(\tau)d\tau\right)+k_2\vect e^\top\int_{0}^{t}\vect e(\tau)d\tau\nonumber\\
    &+\vect\sigma^\top\left(\vect f(\vect x)+\vect u+ \tilde{\vect d}+k_1\dot{\vect e}+k_2\vect e-\ddot{\vect\eta}_d\right)=-k_1\vect e^\top\vect e\nonumber\\
    &+\vect\sigma^\top\left(\vect f(\vect x)+\vect u+ \tilde{\vect d}+k_1\dot{\vect e}+(I+k_2)\vect e-\ddot{\vect\eta}_d\right)
\end{align}
Then, the above Lyapunov function can be negative definite where the following condition is satisfied:
\begin{align}
    \vect f(\vect x)+\vect u+ \tilde{\vect d}+k_1\dot{\vect e}+(I+k_2)\vect e-\ddot{\vect\eta}_d=-k_3\vect\sigma
\end{align}
where $k_3$ is a positive diagonal matrix. Therefore, the equivalent control input can be obtained as follows:
\begin{align}
    \vect\tau_e=M(\vect\eta)\left[-\vect f(\vect\eta)-k_1\dot{\vect e}-(I+k_2)\vect e-k_3\vect\sigma+\ddot{\vect\eta}_d\right]
\end{align}
Then, the Backstepping-based Integral Sliding Mode Control (BS-ISMC) law is designed as:
\begin{align}\label{eq:BS-ISMC}
    \vect\tau=\vect\tau_e-M(\vect\eta)\Gamma \text{sgn}(\vect\sigma)
\end{align}
where $\Gamma$ is a positive diagonal matrix and the second part of the above control law reads as a reaching mode in the context of sliding mode control algorithms.
\subsection{BS-ISMC with Time Delay Estimation}
According to the TDE concept as a part of Time Delay Control (TDC) approach \cite{NEJATBAKHSHESFAHANI201597}, we first introduce a constant diagonal inertia matrix $\bar M$. Then, the control law \eqref{eq:BS-ISMC} can be modified as follows:
\begin{align}\label{eq:law1}
    \vect\tau=&\bar M\left[-k_1\dot{\vect e}-(I+k_2)\vect e-k_3\vect\sigma+\ddot{\vect\eta}_d-\Gamma\text{sgn}(\vect\sigma)\right]\nonumber\\
    &+(C_n+D_n+\text{g}_n)\dot{\vect\eta}+\bar{\vect\tau}_d-\delta
\end{align}
where the nominal model and perturbation model are described as $N=(C_n+D_n+\text{g}_n)\dot{\vect\eta}$ and $\vect p=\bar{\vect\tau}_d-\delta$, respectively.
Let us define an estimation based on the TDE concept for the applied perturbations including the model uncertainties and external disturbances as follows \cite{ESFAHANI2019106526}: 
\begin{align}\label{eq:pertu}
    \tilde {\vect p}=\vect p_{t-L}=\vect\tau(t-L)-N(t-L)-\bar M\ddot{\vect\eta}(t-L)
\end{align}
where $L$ reads as a small delay time for the TDE part and,
\begin{align}
    \ddot{\vect\eta}(t-L)=\frac{\vect\eta(t)-2\vect\eta(t-L)+\vect\eta(t-2L)}{L^2}
\end{align}
Substituting \eqref{eq:pertu} to \eqref{eq:law1}, the proposed BS-ISMC-TDE control law is designed as follows:
\begin{align}\label{eq:Control_Law}
        \vect\tau=&\vect\tau(t-L)+\bar M\left[-k_1\dot{\vect e}-(I+k_2)\vect e-k_3\vect\sigma+\ddot{\vect\eta}_d\right]\nonumber\\
        &+\bar M\left[-\ddot{\vect\eta}(t-L)-\Gamma\text{sgn}(\vect\sigma)\right]+N-N(t-L)
\end{align}
\subsection{Stability Analysis}
A closed-loop system for \eqref{eq:model} with the proposed BS-ISMC-TDE control scheme can be  uniformly ultimately bounded (UUB)
\Lemma {The TDE error $\epsilon_i=\vect p_i-\tilde{\vect p}_i$ is bounded $\epsilon_i^+\geq\left \|\vect p_i-\tilde{\vect p}_i\right \|_\infty$ if the following condition is satisfied:
\begin{align}
    \left \|I-M^{-1}(\vect\eta)\bar M\right \|<1
\end{align}}
In practice, $\bar M$ is adjusted by increasing its diagonal
elements from a small positive value to a large value before the system approaches the oscillatory response. Let us obtain the first-time derivative of sliding surface \eqref{eq:surface} as follows:
\begin{align}
    \dot{\vect\sigma}=\bar M^{-1}\left(\bar{\vect\tau}+\tilde{\vect p}+\epsilon-N\right)+k_1\dot{\vect e}+k_2\vect e-\ddot{\vect\eta}_d
\end{align}
Substituting \eqref{eq:BS-ISMC} to $\bar{\vect\tau}$ in the above equation, the closed loop dynamics can be expressed as:
\begin{align}
    \dot{\vect\sigma}=\bar M^{-1}\epsilon-\Gamma sgn(\vect\sigma)-k_3\vect\sigma
\end{align}
Then, the first-derivative of the Lyapunov function \eqref{eq:V} considering the term including the sliding surface is expressed as follows:
\begin{align}\label{eq:sdot}
    \vect\sigma^\top\dot{\vect\sigma}&=-k_3\vect\sigma^\top\vect\sigma+\vect\sigma^\top\left(\bar M^{-1}\epsilon-\Gamma sgn(\vect s)\right)\nonumber\\
    &=-k_3\vect\sigma^\top\vect\sigma-\Gamma|\vect\sigma|+\vect\sigma^\top\bar M^{-1}\epsilon\nonumber\\
    &\leq-\left(k_3+\Gamma\right)|\vect\sigma|+\vect\sigma^\top\bar M^{-1}\epsilon
\end{align}
Thus, if ${k_3}_{ii}+\Gamma_{ii}>\bar M_{ii}^{-1}\epsilon_i^+$, then $\dot V<0$. However, a proper adjustment of the constant matrix $\bar M$ is a crucial issue to guarantee a stable performance for a trajectory tracking problem. In this paper, we propose an adaptive mechanism to tune the diagonal matrix $\bar M$. Let us define a Lyapunov function as follows:
\begin{align}\label{eq:Vbar}
    \bar V=\frac{1}{2}\vect\sigma^\top\vect\sigma+\frac{1}{2}\sum_{i=1}^{n}\left(\bar M_{ii}\right)^2
\end{align}
Considering \eqref{eq:sdot}, the first derivative of the above Lyapunov function can be defined by the following inequality:
\begin{align}
  \dot{\bar V}&=\vect\sigma^\top\dot{\vect\sigma}+\sum_{i=1}^{n}\bar M_{ii}\dot{\bar M}_{ii}\leq-k_3|\vect\sigma|+\nonumber\\
  &\sum_{i=1}^{n}-\Gamma_{ii}|\vect\sigma_i|+\alpha_i\vect\sigma_i\bar M_{ii}^{-1}+\bar M_{ii}\dot{\bar M}_{ii}
\end{align}
where $\alpha_i$ is a positive scalar value. Hence, the closed loop system is stable if the following equality as dynamics for $\bar M_{ii}$ hold:
\begin{align}\label{eq:Mbar}
      \dot{\bar M}_{ii}=\Gamma_{ii}|\vect\sigma_i|-\alpha_i\left(\frac{\vect\sigma_i}{\bar M_{ii}^2}\right)
\end{align}
\section{Computer Simulations}
\begin{figure}[htbp!]
		\centering
		\includegraphics[width=.7\linewidth]{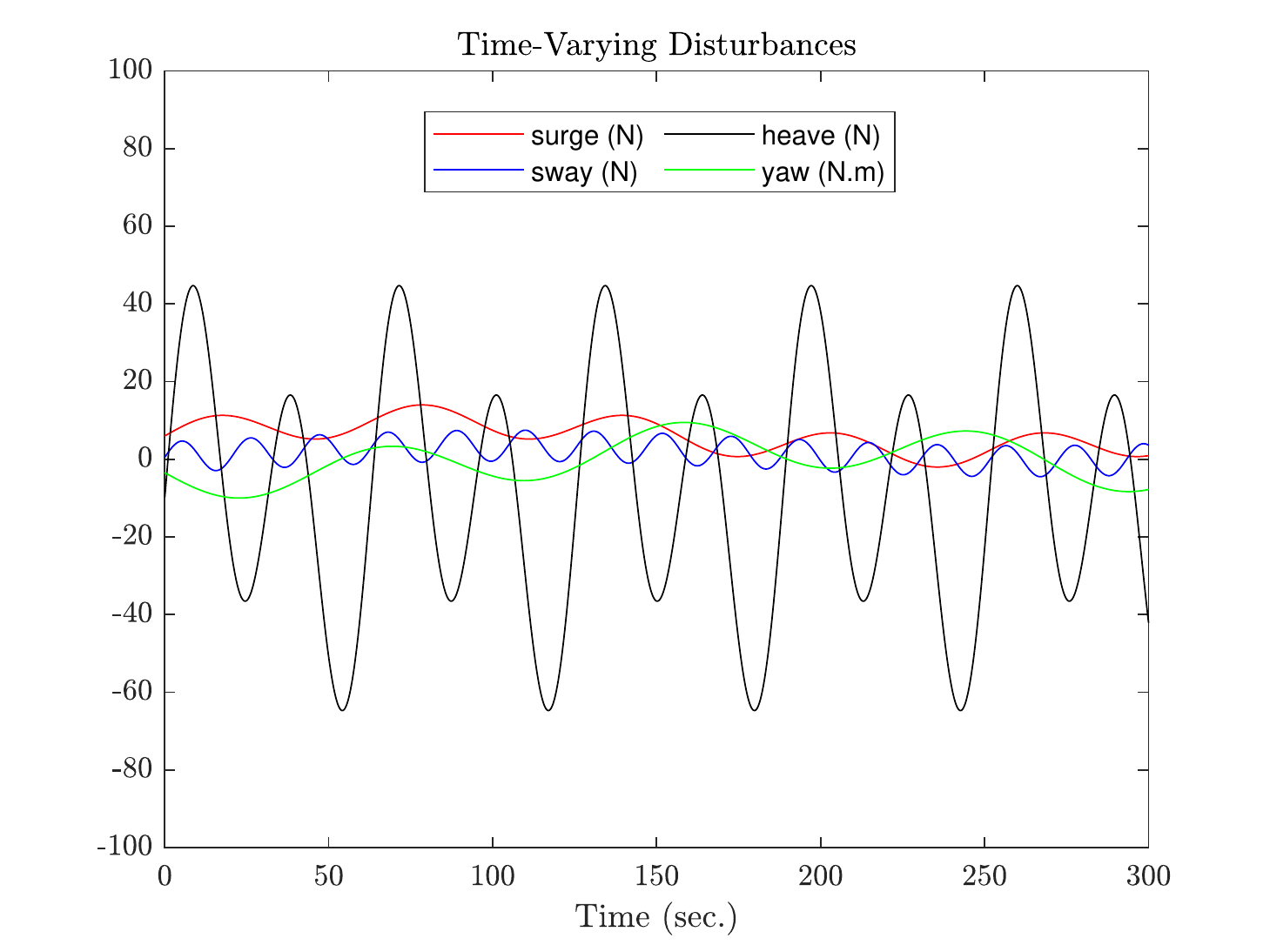}
		\caption{Disturbances applied to surge, sway, heave and yaw motions} 
		\label{dist}
\end{figure}

The numerical values of the AUV parameters used in both examples are given by: $m=54.54kg$, $G=535N$, $B=53.4$, $I_z=13.587$, $X_{uu}=2.3e-2$, $Y_{vv}=5.3e-2$, $Z_{ww}=1.7e-1$, $N_{rr}=2.9e-3$, $X_{\dot u}=-7.6e-3$, $Y_{\dot v}=-5.5e-2$, $Z_{\dot w}=-2.4e-1$, $N_{\dot r}=-3.4e-3$, $X_u=2e-3$, $Y_v=-1e-1$, $Z_w=-3e-1$, $N_r=3e-2$. The parameters of the proposed controllers are given by: $L=T_s=7e-3s$, $\phi=0.2$, $\Gamma=diag\left\{1,1,1,1\right\}$, $k_1=diag\left\{0.1,0.1,0.1,0.1\right\}$, $k_2=diag\left\{0.1,0.1,0.1,0.1\right\}$, $k_3=diag\left\{10,10,10,10\right\}$, $\bar M=diag\left\{3,3,6,1\right\}$.
To alleviate the chattering induced by the switching part in \eqref{eq:Control_Law}, we propose to use the following saturation function: 
\begin{align}
    sat(\vect\sigma)=\left\{\begin{matrix}
sgn(\vect\sigma)\quad &if \quad|\vect\sigma|\geq\phi\\ 
\vect\sigma\quad &if \quad\vect\sigma<\phi
\end{matrix}\right.
\end{align}
where $\phi$ reads as the boundary layer thickness. In both case studies, we apply the disturbances shown in Figure \ref{dist}.
\par \textbf{\textit{Case Study 1}}:
In this case study, the trajectories are generated as follows:
\begin{subequations}
  \begin{align}
    &x=4\text{sin}(0.04t)\\
    &y=2.5\text{cos}(0.02t)\\
    &z=2\text{sin}(0.01t)+2\text{cos}(0.02t)\\
    &\psi=0.5\text{cos}(0.01t)-0.5\text{sin}(0.01t)
  \end{align}
\end{subequations}
where initial conditions are set to $\vect x_0=\left[0,1,2,\frac{\pi}{4}\right]^\top$.
\begin{figure}[htbp!]
		\centering
		\includegraphics[width=1\linewidth]{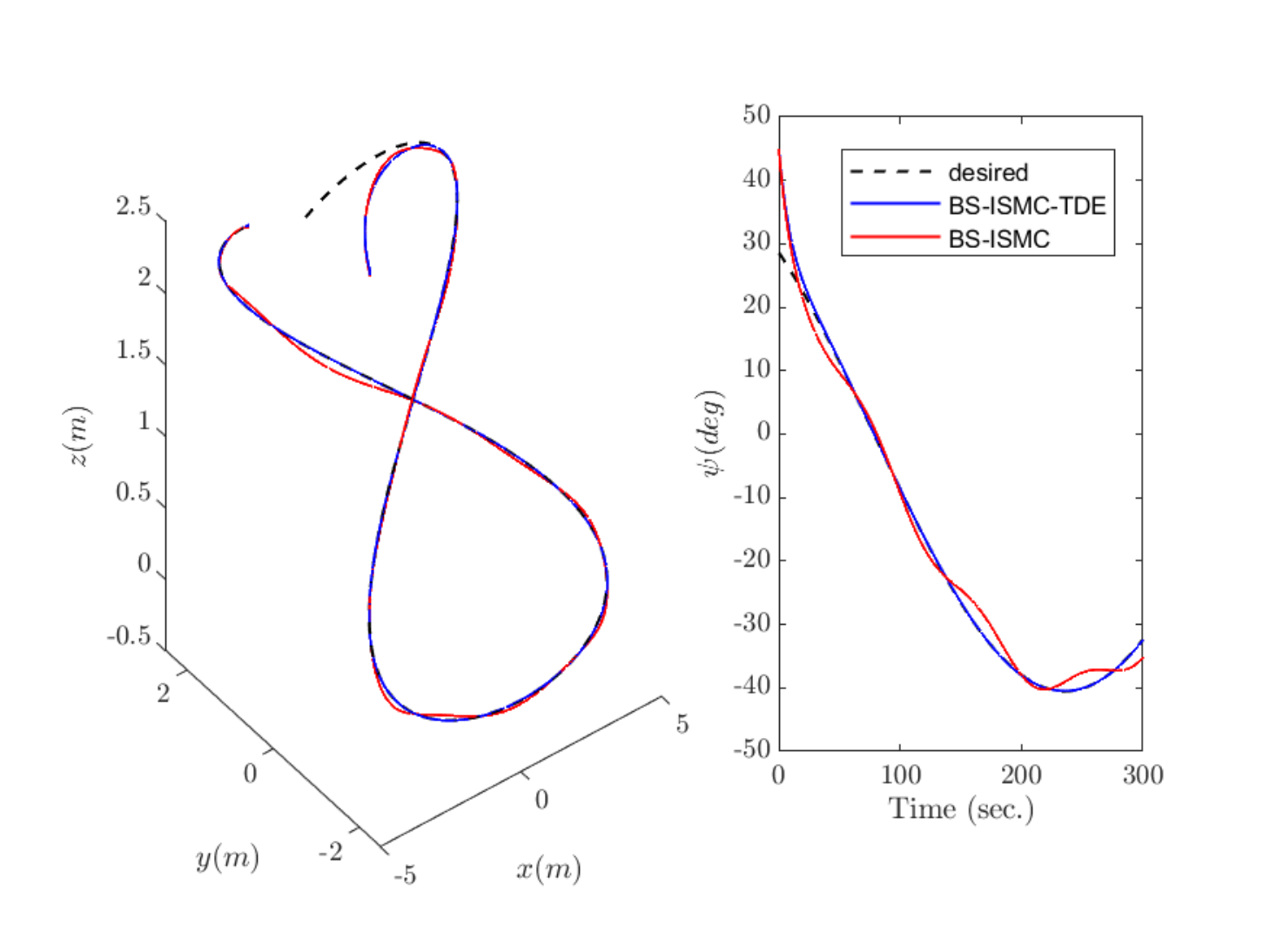}
		\caption{Trajectory tracking control of AUV} 
		\label{traj1}
\end{figure}
\begin{figure}[htbp!]
		\centering
		\includegraphics[width=1\linewidth]{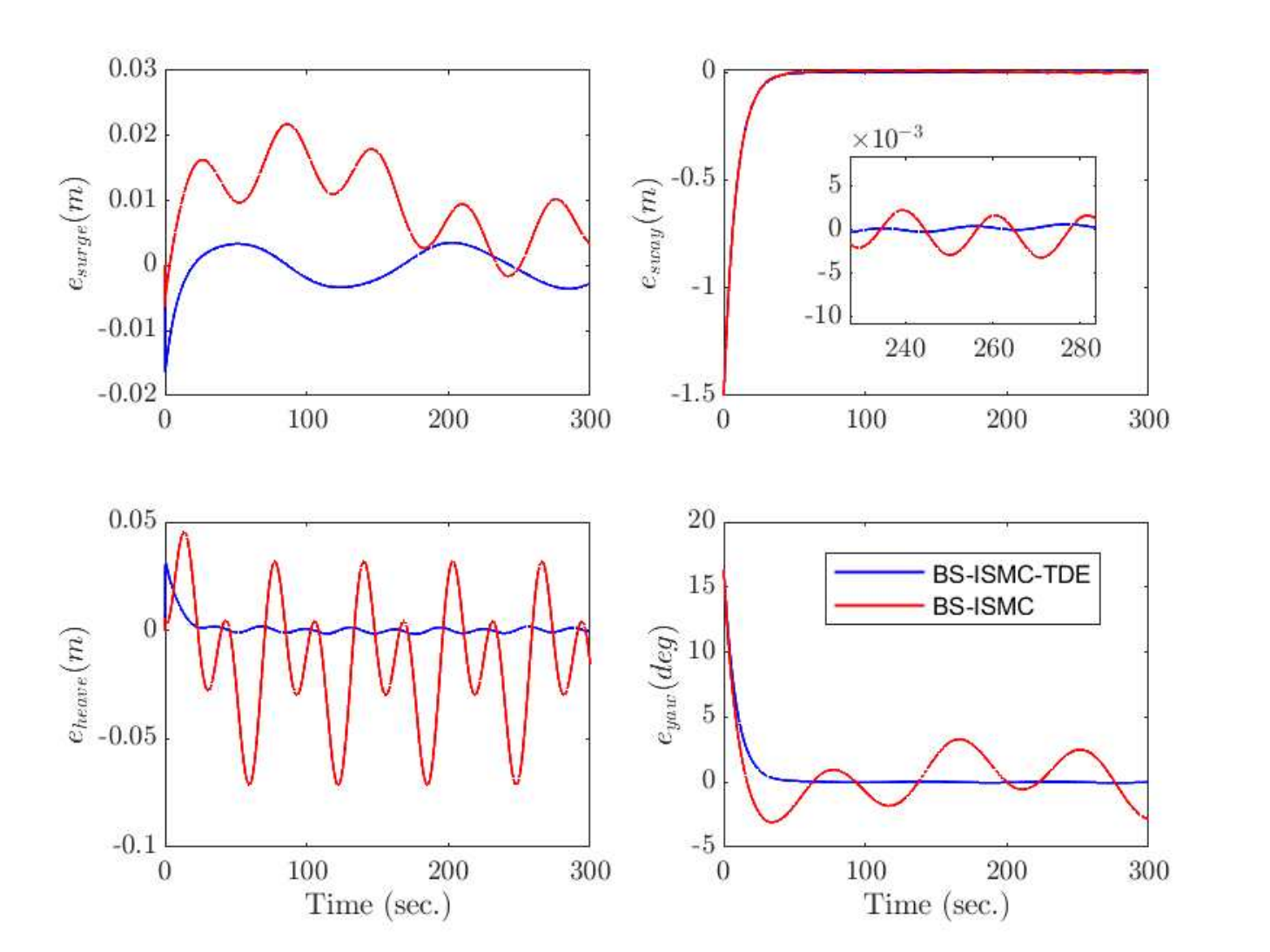}
		\caption{Tracking error signals} 
		\label{error1}
\end{figure}
\begin{figure}[htbp!]
		\centering
		\includegraphics[width=1\linewidth]{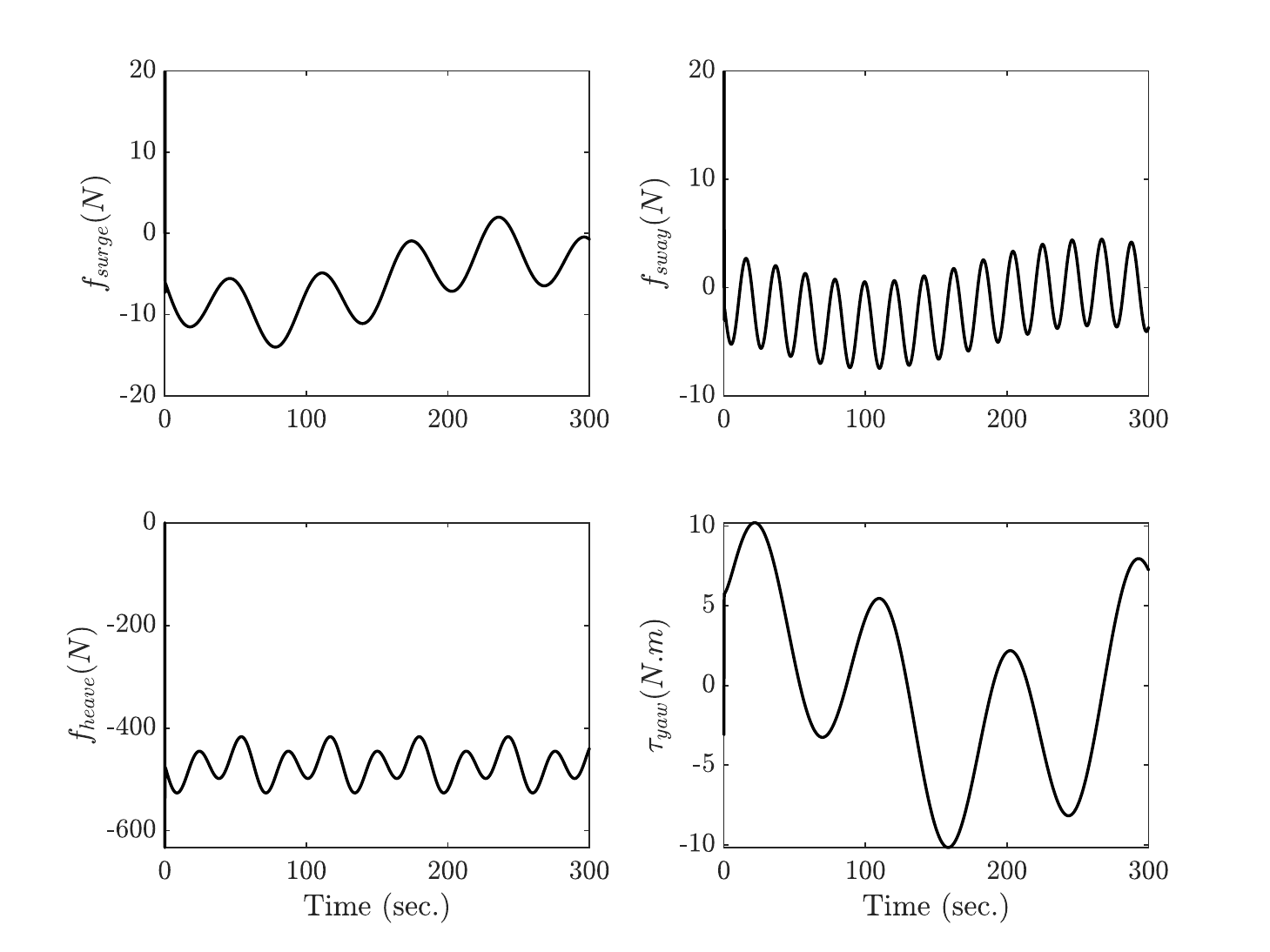}
		\caption{Control inputs for the proposed BS-ISMC-TDE} 
		\label{cont1}
\end{figure}
\begin{figure}[htbp!]
		\centering
		\includegraphics[width=1\linewidth]{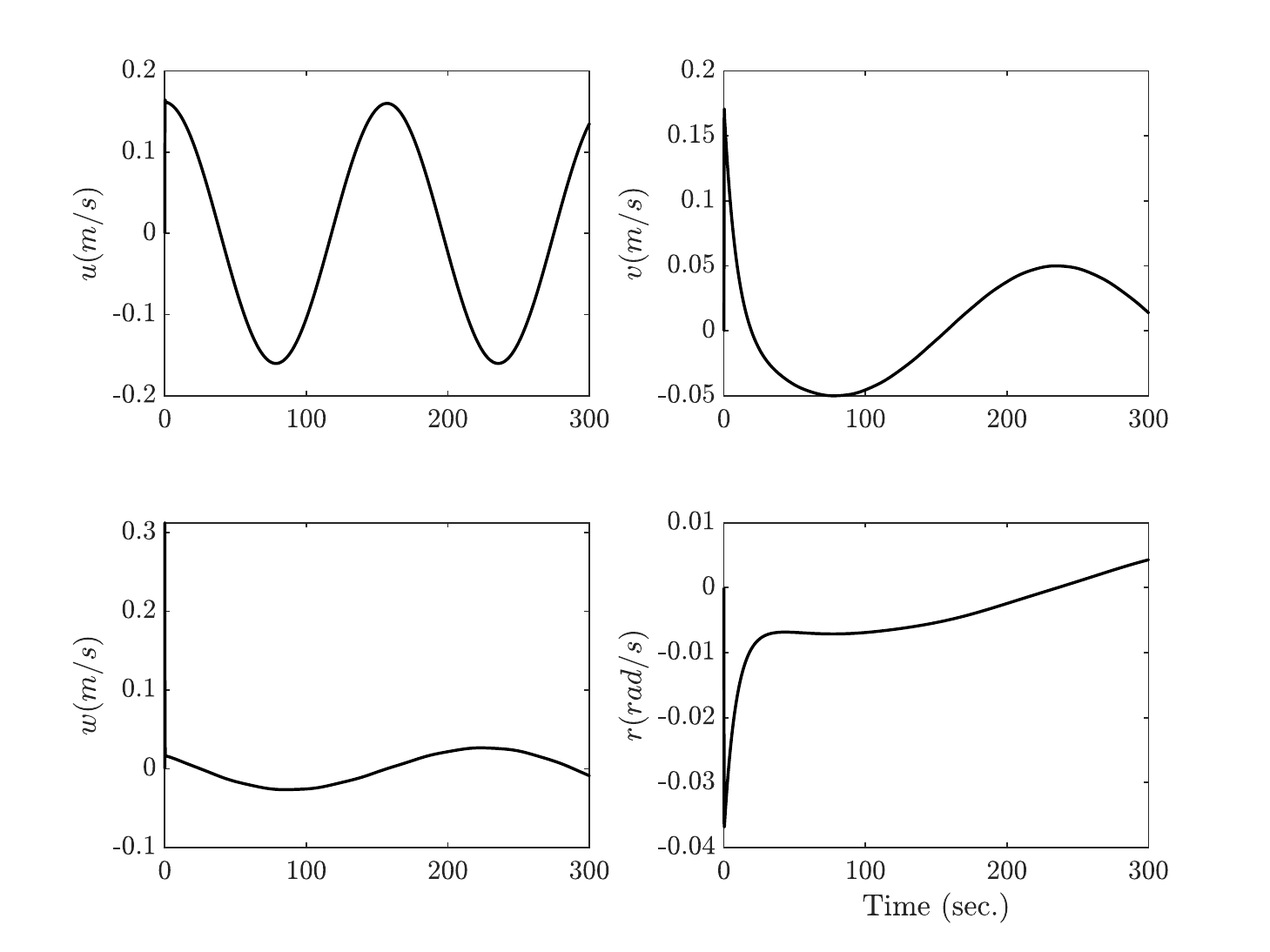}
		\caption{Velocity signals} 
		\label{vel1}
\end{figure}
The differences between the BS-ISMC and a BS-ISMC-TDE in terms of trajectory tracking performance and the tracking errors are illustrated in Figures \ref{traj1} and \ref{error1}. 
\par
\begin{figure}[htbp!]
		\centering
		\includegraphics[width=1\linewidth]{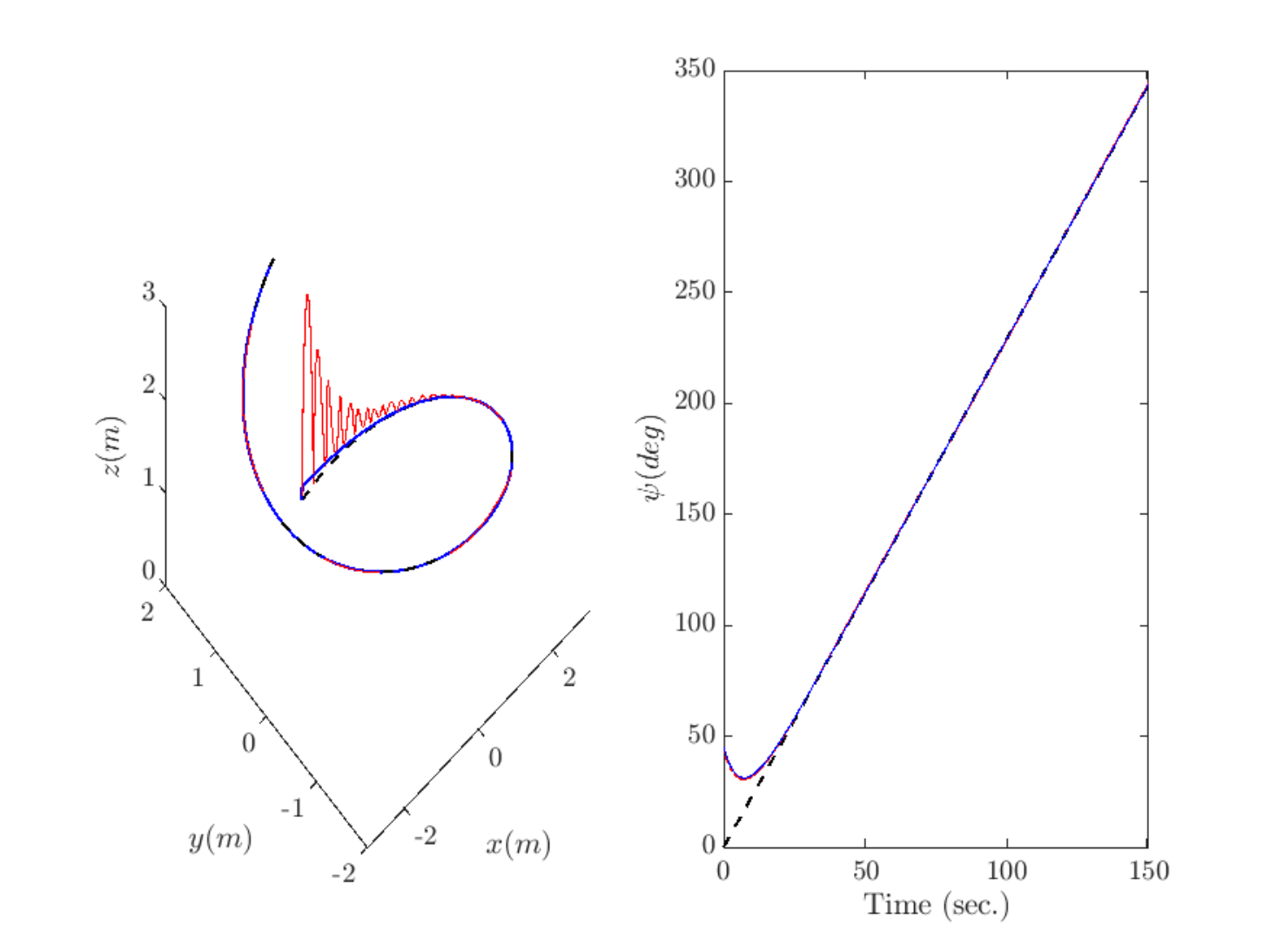}
		\caption{Trajectory tracking control of AUV} 
		\label{2traj1}
\end{figure}

\begin{figure}[htbp!]
		\centering
		\includegraphics[width=1\linewidth]{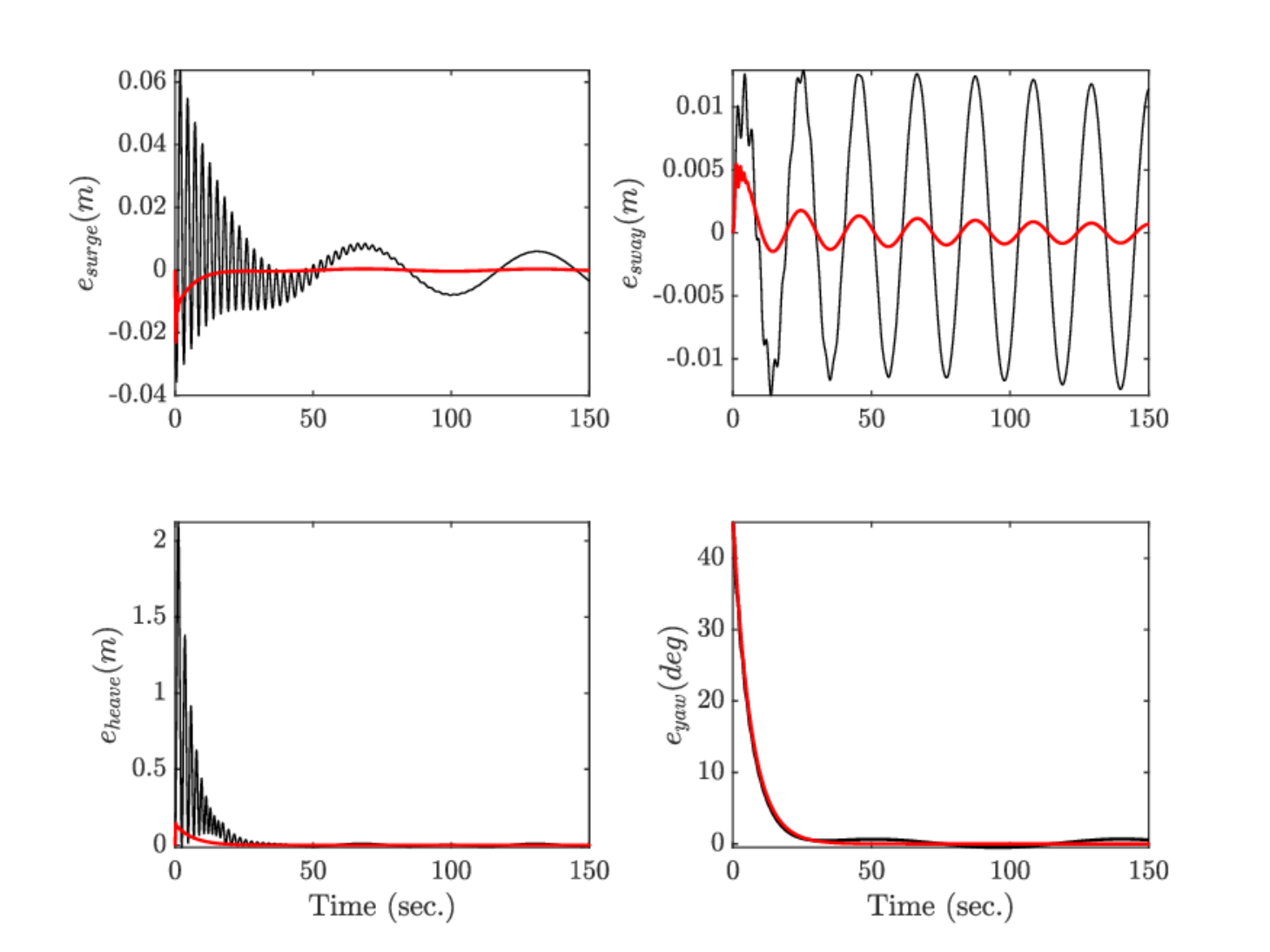}
		\caption{Tracking error signals} 
		\label{2error1}
\end{figure}

\begin{figure}[htbp!]
		\centering
		\includegraphics[width=1\linewidth]{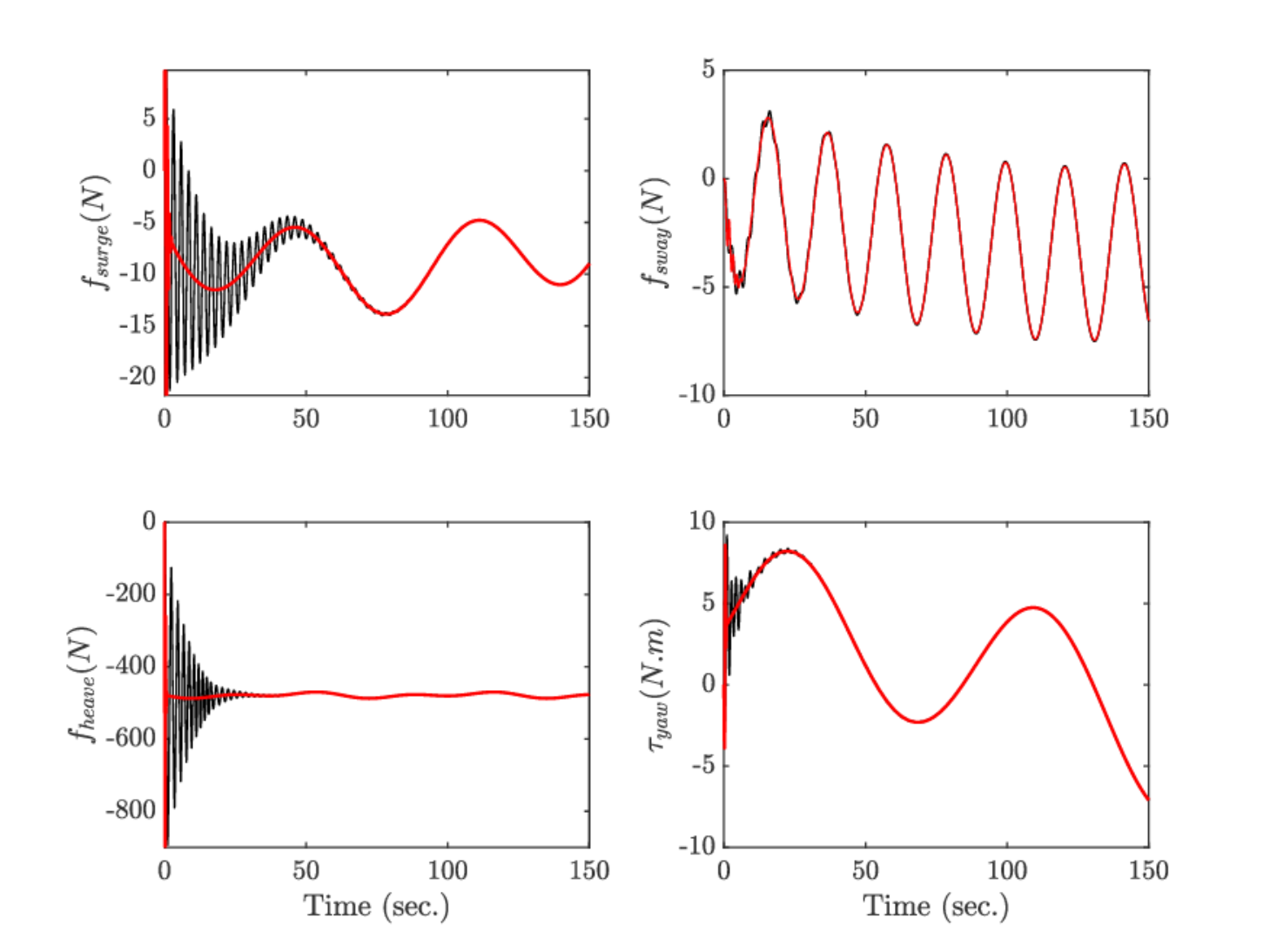}
		\caption{Control inputs for the proposed BS-ISMC-TDE} 
		\label{2cont1}
\end{figure}

\begin{figure}[htbp!]
		\centering
		\includegraphics[width=1\linewidth]{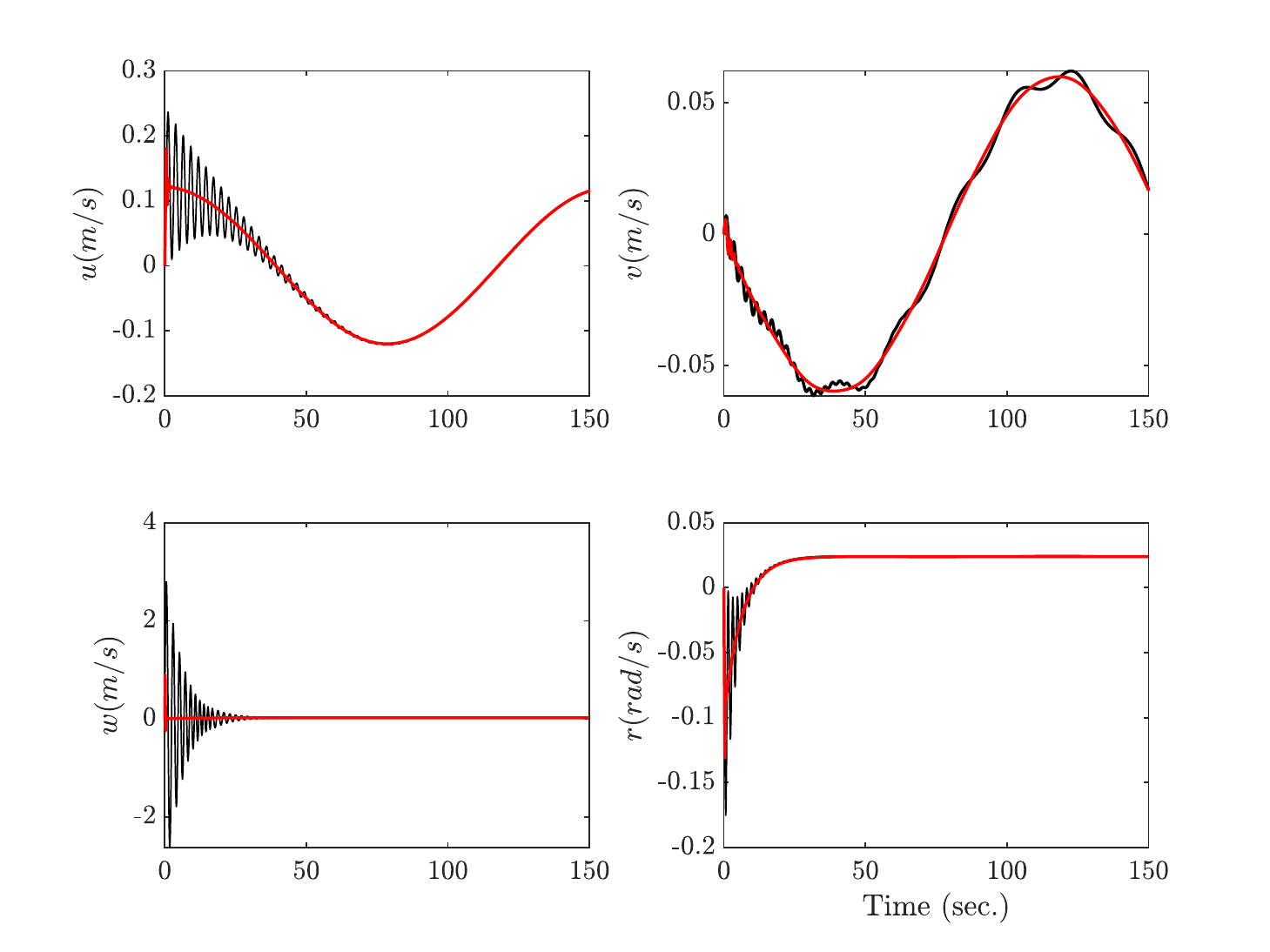}
		\caption{Velocity signals} 
		\label{2vel1}
\end{figure}

\begin{figure}[htbp!]
		\centering
		\includegraphics[width=.93\linewidth]{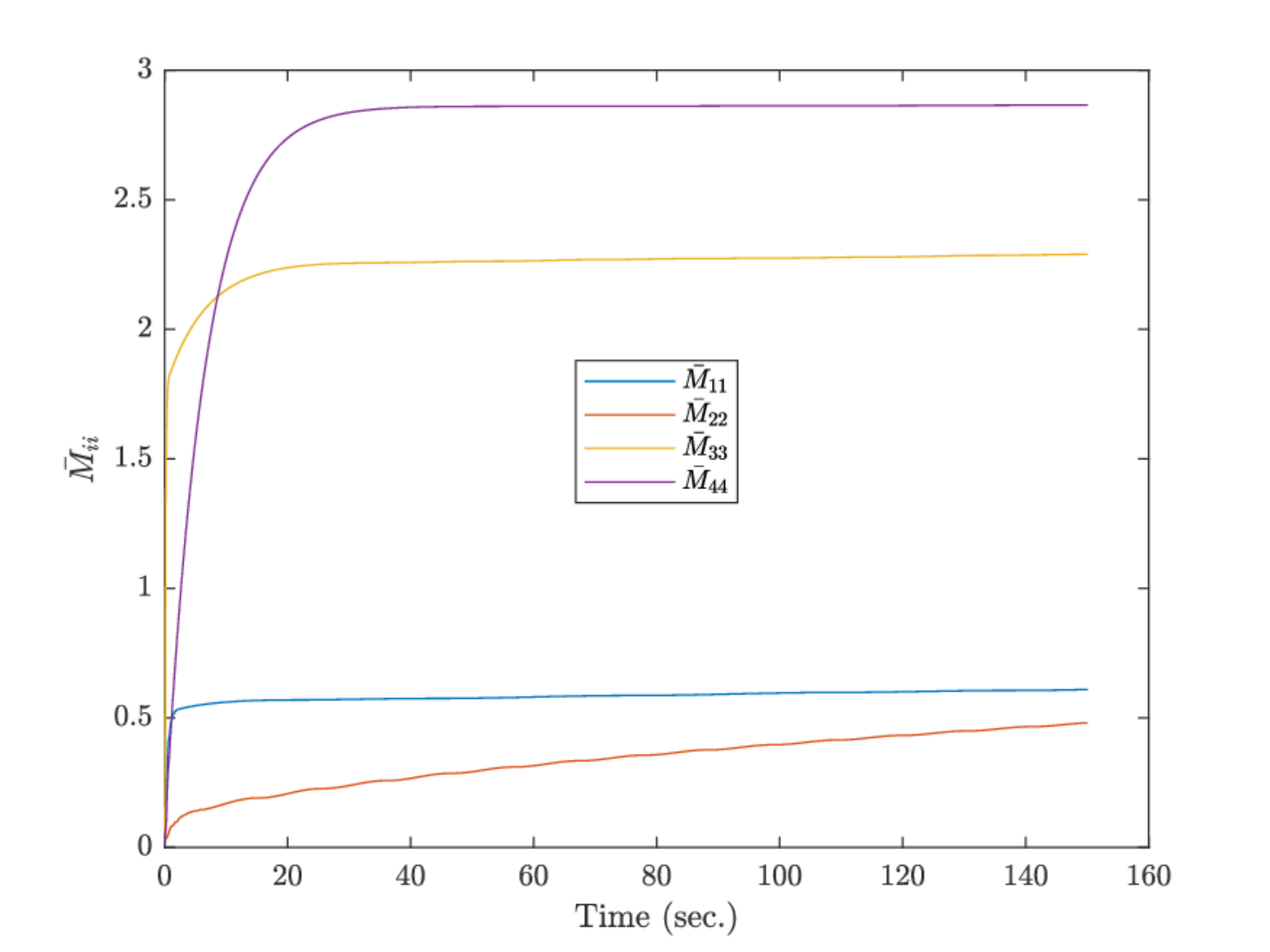}
		\caption{Adjustable matrix $\bar M_{ii}$} 
		\label{mbar}
\end{figure}

 \textbf{\textit{Case Study 2}}:
In this case, a comparative study on the proposed BS-ISMC-TDE is accomplished. To investigate the effectiveness of the proposed adaptive TDE with adjustable $\bar M$ shown in Figure \ref{mbar}, we initialize the values of diagonal matrix with $\bar M_{ii}=diag\left([0.03,0.03,0.05,0.02]\right)$ so that the response is very close to the instability and then let the adaptive mechanism to tune this matrix. The initial conditions are set to $\vect x_0=\left[0,1.5,0,\frac{\pi}{4}\right]^\top$. The sampling time $T_s$ is equal to the delay time $L=2ms$. The  reference trajectories  are generated as follows:
\begin{subequations}
  \begin{align}
    &x=3\text{sin}(0.04t)\\
    &y=1.5\text{cos}(0.04t)\\
    &z=0.02t\\
    &\psi=0.04t
  \end{align}
\end{subequations}
As it is observed, the proposed BS-ISMC-TDE with an adaptive gain part can rapidly bring a poorly-adjusted $\bar M$ with an aggressive response of the controller to a well-adjusted and a smooth and stable behavior shown in Figures \ref{2traj1} to \ref{2vel1}.

\section{Conclusion}
In this paper, we proposed to leverage three robust control schemes for improving a trajectory tracking control of autonomous underwater vehicles in presence of external time-varying disturbances and uncertainties. In the proposed ,Backstepping-based Integral Sliding Mode Control with Time Delay Estimation (BS-ISMC-TDE), the backstepping part is responsible for tracking the desired trajectory while the  Integral Sliding Mode Control (ISMC) is designed for tackling the uncertainties and external time varying disturbances. To improve the accuracy of the tracking in presence of some unknown parameters and to compensate the effect of unmodeled dynamics, the Time Delay Estimation (TDE) is applied to the BS-ISMC. Besides, an adaptive mechanism for adjusting the TDE part is proposed to address a problem with a poorly-adjusted $\bar M$. As a future work, we will investigate a fully adaptive BS-ISMC-TDE in which all gains related to the BS and ISMC parts are autonomously adjusted as well.
	\bibliographystyle{IEEEtran}
	\bibliography{IEEEabrv,main}
\end{document}